\documentstyle[aps]{revtex}
\begin{document} 
%\draft 
\title{Selective Electrodeposition of Nanometerscale
Magnetic Wires} 
\author{Gerhard Fasol\footnote{Permanent address: Eurotechnology
Japan K. K., Parkwest Building, 6-12-1 Nishi-Shinjuku, Shinjuku-ku, Tokyo 160,
Japan; Email: g.fasol@ieee.org; WWW: http://www.euro-technology.com}}
\address{`Sakigake 21'-PRESTO, Japan Science and Technology Corporation (JST),
and Institute of Industrial Science, University of Tokyo, 7-22-1 Roppongi,
Minato-ku, Tokyo 106, Japan} 
\author{Katharina Runge\footnote{Permanent address:
Intocast Japan Corp., 2-12-1 Shinbashi, Minato-ku, Tokyo 105, Japan, Email:
katharin@gol.com}} 
\address{Institute of Industrial Science, University of Tokyo,
7-22-1 Roppongi, Minato-ku, Tokyo 106, Japan}

\maketitle

\begin{abstract} 
A selective electrodeposition method for the fabrication  of
extremely thin and long metallic and magnetic wires and other nanostructures is
introduced. Growth is done on the cleaved edge of a semiconductor multilayer
structure incorporating a 4~nm wide modulation doped quantum well. This
conducting quantum well is connected to the negative  current contact during
electrodeposition. Since electrodeposition requires the neutralization of
positive metal ions from the solution, deposition takes place selectively onto
the  edge of the quantum well, leading to the fabrication of extremely thin
magnetic metal wires, which should be useful for the investigation of the limits
of magnetic storage. 
\end{abstract}

\pacs{75.50.Ss, 81.15.Pq}

Extremely thin magnetic and non-magnetic metal  wires are of great fundamental
and practical interest. The study of transport properties of extremely thin metal
wires is important to clarify localization phenomena, while nanometer scale
magnetic wires are of fundamental and technological relevance for the exploration
of the extreme limits of magnetic storage\cite{wernsdorfer}. However, the
precisely controlled fabrication of smaller and smaller nanometerscale wires is
extremely challenging.

Electrodeposition of ultrafine wires using templates has been first used for the
fabrication of nanowires by Possin\cite{possin}. Possin fabricated short Sn, In,
and Zn nanowires with diameters of 30 nm and   lengths of up to 15 $\mu$m by
electrodeposition into pores fabricated by etching of the damage produced by high
energy charged particles into mica\cite{price}. Nanowires have been deposited
also into the pores of electrochemically oxidized Aluminium\cite{kawai}, and into
pores etched into plastic membranes\cite{penner}. Growth of magnetic wires into
pores has been reported by AlMawlawi\cite{almawlawi}. Electrochemical fabrication
of magnetic structures has been recently reviewed by Schwarzacher and
Lashmore\cite{schwarzacher}. Ultranarrow but very short structures have also been
fabricated by a scanning tunneling microscope (STM)\cite{eigler}.

In the present work we introduce a new electrodeposition method onto a layered
semiconductor template structure to fabricate nanoscale metal and magnetic wires.
We report magnetic wires with a width of 20~nm, however we expect that our method
should be applicable down to widths of around 4~nm, and to a variety of
differently shaped structures, including closed squares in addition to straight
wires. Our method greatly expands the variety of metallic and magnetic
nanostructures accessible to fabrication.

The advantage of the method  is that it allows the fabrication of ultrafine wires
down to 4~nanometers with much more control over the shapes than is possible with
growth into etched nanopores. Also it allows the controlled deposition of several
millimeter long parallel nanowires onto semiconductor substrates. In addition,
much longer wires can be produced than is possible in the fairly narrow range of
view of an STM.

A schematic view of our fabrication method is shown in Figure~\ref{method}. The
basic process is electrodeposition where positively charged metal ions from a
solution recombine with electrons delivered by an electrode to form neutral
deposited atoms onto the same electrode. We use the edge of a conducting
4~nanometer thin InAs quantum well, embedded into a semiconductor multilayer
structure, as the electrode. Thus the 4~nm wide edge of the InAs conducting
quantum well acts as a template for selective electrodeposition. We use an
InAs-edge because  the Fermi level on the surface of InAs is pinned within the
conduction band. Thus the surface depletion layer customary in GaAs-based
materials is avoided.  The InAs well therefore forms an extremely thin conducting
layer of which one edge is exposed to the electrolyte at the cleaved edge. Here
we report results on Permalloy as the deposited material, but a range of other
materials are under investigation.

The following sequence of layers is grown by molecular beam epitaxy (MBE) onto an
InP substrate: a  200~nm undoped
$\mathrm{In}_{0.52}\mathrm{Al}_{0.48}\mathrm{As}$ buffer layer (200~nm), a 13~nm 
$\mathrm{n}^{+}$ Si-doped ($4 \times 10^{18} \mathrm{cm}^{-3}$)
$\mathrm{In}_{0.52}\mathrm{Al}_{0.48}\mathrm{As}$, a 6~nm undoped
$\mathrm{In}_{0.52}\mathrm{Al}_{0.48}\mathrm{As}$ spacer layer, a 4~nm undoped
$\mathrm{In}_{0.53}\mathrm{Ga}_{0.47}\mathrm{As}$ spacer layer, a 4~nm undoped
$\mathrm{In}\mathrm{As}$ modulation doped channel, a 12~nm undoped
$\mathrm{In}_{0.53}\mathrm{Ga}_{0.47}\mathrm{As}$ spacer layer, a 20~nm undoped
$\mathrm{In}_{0.52}\mathrm{Al}_{0.48}\mathrm{As}$ spacer layer and finally a 2~nm
undoped $\mathrm{In}_{0.53}\mathrm{Ga}_{0.47}\mathrm{As}$ protective layer.

An approximately $ 10 \mathrm{mm} \times  15 \mathrm{mm}$ large chip was cleaved
from the wafer. The conducting modulation doped layer was contacted at one point
at one end of the cleaved chip (opposite to the end where the wire was deposited)
by thermally diffusing Indium under a controlled Argon atmosphere in a reaction
chamber. A thin gold wire was soldered onto the diffused Indium contact. The
contacted chip was then suspended on the Gold wire into the electroplating bath.
For electro-deposition this contact was connected to the negative output of a
constant current source. A  Nickel wire was connected to the positive output of
the constant current source and inserted into a citrate-complexed nickel-iron
electrolyte\cite{venkatasetty} with a pH of around 4.5. The container of the
plating bath was temperature stabilized to $(29.2 \pm 0.5)^{\circ}{\mathrm C}$.
Typically currents of $100~\mu\mathrm{A}$ were used. The fabricated metal wires
were investigated using an atomic force microscope (AFM). Figure~\ref{afm} shows
an AFM image of the top view of the cleavage plane of the wafer with a 20~nm
permalloy wire. Starting from the left hand side, first the edge of the wafer is
seen, and then close to it the permalloy wire. In addition to the  wire, several
approximately 20~nm large spherical particles are also detected. The origin of
these particles is unknown. The contrast between  the 261~nm MBE deposited
multilayer structure and the substrate can also be seen on the AFM image.

In conclusion, we have introduced and demonstrated a new fabrication method for
the selective deposition of extremely thin metallic wires. We have demonstrated
the deposition of a 20~nm permalloy wire. The present method should also allow
the deposition of wires down to widths of about 4~nm, enabling the fabrication of
a rich variety of structures. The present method has many advantages compared to
competing methods: for example it enables the fabrication of millimeter long
wires.  We expect the present method to be of great use for the exploration of
the magnetic properties of  metallic nanostructures, and possibly even for the
fabrication of novel magnetic storage or read/write devices.

One of the authors (KR) would like to thank the European Union for financial
support. This project would have been impossible without the support of the Japan
Science and Technology Corporation (JST) under the `Sakigake 21' Research
Programme---in particular, the help and support by K.  Kohra, H. Tanaka, and Y.
Ogawa is gratefully acknowledged. The authors express their gratitude to
Professor C. Stanley (University of Glasgow) for the growth of the MBE
structures.

\begin{figure} \caption{This schematic drawing shows a new method of selective
electrodeposition using the cleaved edge of a conducting 4 nm InAs modulation
doped quantum well  embedded into a layered semiconductor structure as a
template. This conducting InAs quantum well is connected to the negative  current
contact during electrodeposition, leading to the selective growth of an extremely
thin magnetic metal wire.} \label{method} \end{figure}

\begin{figure} \caption{Atomic force microscope image of a 20~nm thin magnetic
wire deposited onto the cleaved edge of an MBE grown multilayer structure. In
addition to the desired wire structure, several undesired spherical particles can
also be observed. The contrast between the  261~nm thick   MBE multilayer
structure and the substrate can also be seen in the AFM image.} \label{afm}
\end{figure}


\begin{references}

\bibitem{wernsdorfer}W. Wernsdorfer, B. Doudin, D. Mailly, K. Hasselbach, A.
Benoit, J. Meier, J.-Ph. Ansermet, B. Barbara, Phys. Rev. Lett. {\bfseries 77},
1873 (1996).

\bibitem{possin} G. E. Possin, Rev. Sci. Instr. {\bfseries 41}, 772 (1970).

\bibitem{price}P. B. Price and R. M. Walker, J. Appl. Phys. {\bfseries 33}, 3407
(1962).

\bibitem{kawai}S. Kawai and R. Ueda, J. Electrochem. Soc. {\bfseries 122}, 32
(1975).

\bibitem{penner}R. M. Penner and C. R. Martin, Anal. Chem. {\bfseries 59}, 2625
(1987).

\bibitem{almawlawi}D. AlMawlawi, N. Coombs and M. Moskovits, J. Appl. Phys.
{\bfseries 70}, 4421 (1991).

\bibitem{schwarzacher}W. Schwarzacher and D. S. Lashmore, IEEE Transactions on
Magnetics {\bfseries 32}, 3133 (1996).

\bibitem{eigler}D. M. Eigler and E. K. Schweizer, Nature {\bfseries 344}, 524
(1990).

\bibitem{venkatasetty}H. V. Venkatasetty, J. Electrochem. Soc. {\bfseries 117},
403 (1970).

\end{references}
\end{document}